\documentclass[%
superscriptaddress,
preprint,%
showpacs,%
showkeys,
nobibnotes,
amsmath,amssymb,%
aps,%
pre,%
longbibliography,
floatfix,
]{revtex4-1}

\usepackage[utf8]{inputenc}
\usepackage[T1]{fontenc}
\usepackage[english]{babel}
\usepackage{amsfonts}
\usepackage{amsmath}
\usepackage{graphicx}
\usepackage{color}

\newcommand{\e}{\varepsilon}              %

\allowdisplaybreaks

\begin{document}

\title{Multi-pulse phase resetting curves}

\author{Giri P. Krishnan}
\affiliation{Department of Cell Biology and Neuroscience, University of California, Riverside CA, USA}
\affiliation{Department of Physics and Astronomy, University of Potsdam, 14476
Potsdam, Germany}
\author{Maxim Bazhenov}
\affiliation{Department of Cell Biology and Neuroscience, University of California, Riverside CA, USA}
\author{ Arkady Pikovsky}
\affiliation{Department of Physics and Astronomy, University of Potsdam, 14476
Potsdam, Germany}

\begin{abstract}

In this paper, we introduce and systematically study, in terms of phase response curves (PRC), 
the effect of a dual pulse excitation 
on the dynamics of an autonomous oscillator.  
Specifically, we test the deviations from a linear summation of phase 
advances from two small perturbations. We derive analytically 
a correction term, which generally appears for oscillators whose 
intrinsic dimensionality is greater than one. We 
demonstrate this  effect in case of the Stuart-Landau model,
 and also in various higher  dimensional neuronal model.  
 The non-linear correction is found to be proportional  to the square of 
the perturbation. This deviation from the superposition principle needs to be 
taken  into account in studies of networks of pulse-coupled oscillators. 
Further, this deviation  could be used for verification of oscillator models
via a dual pulse excitation. 
\end{abstract}

\date{\today}
\pacs{05.45.Xt,87.19.ll}
\maketitle

\section{Introduction}
\label{sec:intro}

The weakly connected oscillator theory is often used to show conditions under  
which the oscillators are synchronized~\cite{Kuramoto-84}. 
In this theory, small perturbations do not influence the amplitude of the 
oscillator, however they have a significant effect on its phase. 
This allows one for a drastic reduction in the description of the oscillator: 
instead of operating with the original, possibly  high-
dimensional set of equations, only one phase variable is used for each 
oscillator. Using the phase description alone, it is possible to find the 
conditions when oscillators are synchronized, provided the phase reduction 
is accurate enough, what is the case  if the coupling is weak. One of 
the assumptions used in this  approach is the principle of superposition, 
which states that the effect of several  small perturbations on the 
period of the oscillation could be considered independently, and then summed.  

In this paper, we examined the phase dynamics beyond the superposition 
principle. More precisely, we consider the effect of two relatively small 
perturbations  on the phase for various types of oscillators.  Our main 
tool in the description of the phase dynamics is the  Phase Response 
Curve (PRC, $ \delta{\varphi} $) which is widely  used in both theoretical 
and experimental studies, especially in the field of  
neuroscience~\cite{Winfree-80,Kuramoto-84, Glass-Mackey-88, Canavier-06, Schultheiss-Prinz-Butera-12}.  
The PRC measures the shift of the phase 
of the 
oscillator due to an external pulse, in dependence on the phase at which  
the pulse is applied. Or, in terms of the oscillator period, the PRC 
measures the local change in the period of the oscillator due to a 
pulse perturbation at various time points within the period. The oscillation can 
either advance or delay based on the sign of the PRC.  To determine the PRC, 
one often performs an experiment just according to the definition 
of the PRC; this can be accomplished for individual biological 
neurons~\cite{Galan-Ermentrout-Urban-05} and for complex oscillating systems like 
those responsible for circadian rhythms in the 
brain~\cite{Khalsa_etal-03,Hilaire_etal-12}, see 
Refs.~\cite{Phoka_et_al, Ikeda-Yoshizawa-Sato-83, Abramovich-Sivan_Akselrod-98} 
for other biophysical examples. Furthermore, the PRC concept can be 
applied not only to individual 
oscillators, but to collective modes 
as well~\cite{Ko-Ermentrout-09, Levnajic-Pikovsky-10}.  

The shape of the PRC curve is shown to be critical for the synchronization 
properties of the oscillator  networks~\cite{Achuthan-Canavier-09}. 
However, in the synchronization problem, an oscillator is subject not just 
to one external pulse, but to a series of pulses from  the external 
force or another oscillator (or several other oscillators if more then two 
oscillators are coupled). Thus, in order to apply the PRC concept in 
such a situation, one has to know how the oscillator responds to a series 
of pulses. If the superposition principle holds, then the sum of two 
small perturbations will independently influence the period of the oscillator 
according to the PRCs  for single inputs, and thus can be linearly 
added to predict the overall phase shift.  However, generally one expects 
deviation from this simple superposition if the perturbations are not 
small. In this paper we systematically consider the effect of two pulses 
on the oscillator's phase, and characterize the deviations from the 
pure superposition as nonlinear effects. We illustrate these effects also 
for several realistic models of neuron dynamics.

\section{Phase dynamics and definition of multi-pulse PRC}
\label{sec:pd}

\subsection{Pure phase dynamics}
\label{sec:ppd}
We start with the simplest case where the oscillator is described just by one variable, the phase
$\varphi$ (to be assumed $2\pi$-periodic) that grows uniformly in time
\begin{equation}
\dot\varphi=\omega
\label{eq:phdyn}
\end{equation}
Suppose that the action of a forcing pulse with strength $\e$ is described by 
the standard PRC $\e S(\varphi,\e)$ (here dependence of $S$ on $\e$ accounts 
for nonlinear terms, so that
$S(\varphi,0)$ is the linear PRC). Consider now the action of two pulses, at 
times $t_0$ and $t_0+\tau$, having strengths $\e_0$ and $\e_1$, respectively. 
Just after the first pulse
\[
\varphi_+(t_0)=\varphi(t_0)+\e_0S(\varphi(t_0),\e_0) \;.
\]
Just prior to the second pulse the phase is 
\[
\varphi(t_0+\tau)=\varphi_+(t_0)+\omega\tau=\varphi(t_0)+\omega\tau+\e_0S(\varphi(t_0),\e_0) \;,
\]
and after the second pulse
\[
\varphi_{+}(t_0+\tau)=\varphi(t_0)+\omega\tau+\e_0S(\varphi(t_0),\e_0)+\e_1 
S(\varphi(t_0)+\omega\tau+\e_0S(\varphi(t_0),\e_0),\e_1)\;.
\]
Thus, the overall effect of two pulses
\begin{equation}
\Delta\varphi=\e_0S(\varphi(t_0),\e_0)+
\e_1S[\varphi(t_0)+\omega\tau+\e_0S(\varphi(t_0),\e_0),\e_1]
\label{eq:genprc0}
\end{equation}
can be simply calculated via the superposition of two single-pulse PRC 
functions $\e S(\varphi,\e)$. In the linear approximation, 
where $\e S(\varphi,\e)\approx \e S(\varphi,0)$ one obtains just the sum
\[
\Delta\varphi\approx \e_0 S(\varphi(t_0),0)+\e_1 S(\varphi(t_0)+\omega\tau,0)\;.
\]

\subsection{General consideration}
\label{sec:gc}

Now we consider a general situation where periodic oscillations are described
by a limit cycle $\mathbf{x}_0(t)$ in a generally $N$-dimensional phase space.
The crucial notion simplifying the consideration, is that of 
isochrons~\cite{Guckenheimer-75}, which are 
submanifolds of codimension one foliating the phase space and having 
the same phase 
as the corresponding points on the limit cycle. This allows one to 
represent the phase space
as $(\mathbf{a},\varphi)$ where $\mathbf{a}$ is 
an $(N-1)$-dimensional ``amplitude'', and the phase
obeys the same equation (\ref{eq:phdyn}). Without loss of generality, 
to simplify notations, 
we can assume that on the limit cycle the amplitude vanishes $\mathbf{a}=0$.

In terms of the phase and the amplitude, a pulse that kicks the 
system resets the 
state  $(\mathbf{a},\varphi)$ as
\begin{align*}
\varphi&\to\varphi+\e\Phi(\mathbf{a},\varphi,\e)\;,\\
\mathbf{a}&\to\mathbf{a}+\e\mathbf{A}(\mathbf{a},\varphi,\e)\;,
\end{align*}
where again $\Phi(\mathbf{a},\varphi,0)$ and $\mathbf{A}(\mathbf{a},\varphi,0)$
correspond to a linear approximation.
The usual PRC is defined for the initial state on the limit cycle ($\mathbf{a}=0$), 
so $S(\varphi,\e)=\Phi(0,\varphi,\e)$.
We now apply as above two pulses at times $t_0$ and $t_0+\tau$, assuming that
the system is initially on the limit cycle. Then after the first pulse
\[
\varphi_+(t_0)=\varphi(t_0)+\e_0 S(\varphi(t_0),\e_0) 
,\qquad \mathbf{a}_+(t_0)=\e_0 \mathbf{A}(0,\varphi(t_0),\e_0)\;.
\]
Just prior to the second pulse
\begin{align*}
\varphi(t_0+\tau)&=\varphi_+(t_0)+\omega\tau=\varphi(t_0)+\omega\tau+\e_0 
S(\varphi(t_0),\e_0),\\
\mathbf{a}(t_0+\tau)&=\mathcal{L}^{\tau}(t_0)\mathbf{a}_+(t_0)=
\mathcal{L}^{\tau}(t_0)\e_0 \mathbf{A}(0,\varphi(t_0),\e_0)\;,
\end{align*}
where $\mathcal{L}^{\tau}$ is the operator describing the evolution 
of the amplitudes.
After the second pulse the new phase is
\[\varphi_+(t_0+\tau)=
\varphi(t_0+\tau)+\e_1\Phi(\mathbf{a}(t_0+\tau),\varphi(t_0+\tau),\e_1)
\]
and the overall phase shift due to two pulses (two-pulse PRC) is
\begin{equation}
\delta \varphi=\e_0 S(\varphi(t_0),\e_0)+\e_1\Phi(\mathbf{a}(t_0+\tau),\varphi(t_0+\tau),\e_1)\;.
\label{eq:genprc1}
\end{equation}
Comparing this with expression (\ref{eq:genprc0}) we see that now the 
effect is not a superposition
of two PRCs, but contains the amplitude-dependent phase reset function $\Phi$. 
The difference between  
expressions  (\ref{eq:genprc0}) and (\ref{eq:genprc1}) gives the nontrivial 
effect of multiple pulses
on the phase of the oscillator as the nonlinear correction term $\Delta$:
 \begin{equation}
 \begin{aligned}
&\Delta=\delta \varphi-\Delta\varphi=\\
&=\e_1 \Phi(\mathcal{L}^{\tau}(t_0)\e_0\mathbf{A}(0,\varphi(t_0),\e_0),
\varphi(t_0)+\omega\tau+S(\varphi(t_0),\e_0),\e_1)\\
&-\e_1 S(\varphi(t_0)+\omega\tau+S(\varphi(t_0),\e_0),\e_1)\\
&=\e_1\Phi(\mathcal{L}^{\tau}(t_0)\e_0\mathbf{A}(0,\varphi(t_0),\e_0),
\varphi(t_0)+\omega\tau+S(\varphi(t_0),\e_0),\e_1)\\
&-\Phi(0,\varphi(t_0)+\omega\tau+S(\varphi(t_0),\e_0),\e_1)\;.
\end{aligned}
 \label{eq:genprc2}
\end{equation}
From this expression one can see that the result essentially depends on 
the action of the amplitude
evolution operator $\mathcal{L}^\tau$: 
if $\mathcal{L}^{\tau}(t_0)\mathbf{A}\approx 0$, then the correction
(\ref{eq:genprc2}) vanishes. Thus, the nontrivial effect of the two-pulse 
excitation of
an oscillator depends crucially on the relation between the time interval 
between the pulses
$\tau$ and the relaxation time of the amplitude $t_a$ (characteristic time scale 
of the amplitude evolution operator $\mathcal{L}$), it is mostly pronounced 
if $\tau\lesssim t_a$. If the amplitude is multidimensional, $t_a$ is the time of most slow decay.

In the leading order in the powers of $\e_0,\e_1$, we can represent 
the nonlinear correction as
 \begin{equation}
\Delta \approx \e_1\e_0 \mathcal{L}_l^{\tau}(t_0)\mathbf{A}(0,\varphi(t_0),0)
\frac{\partial}{\partial \mathbf{a}}\Phi(\mathbf{a},\varphi(t_0)+\omega\tau+S(\varphi(t_0),0),0)|_{\mathbf{a}=0}\;,
 \label{eq:genprc3}
\end{equation}
where $\mathcal{L}_l$ is the linearized evolution operator for the amplitudes, 
which describes their relaxation to zero $\sim\exp[-t/t_a]$.

This can be generalized to $ n $ pulses with amplitudes 
($ \e_0,\ldots \e_{n-1},\e_n$) and different time shifts between 
them $(\tau_0,\ldots,\tau_{n-1})$. 
Then the leading terms will be quadratic ones 
($\sim \e_0\e_1\;,\e_0\e_2\;,\e_1\e_2\;,\ldots $), while also 
higher-order corrections
(e.g. $\sim \e_0\e_1\e_2$) will appear. Mostly important will 
be nonlinear terms including neighboring pulses, because, as argued above, 
the effect decreases with the time interval between the pulses. Another straightforward
generalization is the case where two pulses are different and are described by functions
$\Phi_0,\\mathbf{A}_0,\Phi_1,\mathbf{A}_1$.

\subsection{Example: a Stuart-Landau oscillator}
\label{sec:slo}
\begin{figure}
\centering
\includegraphics[width=0.9\columnwidth]{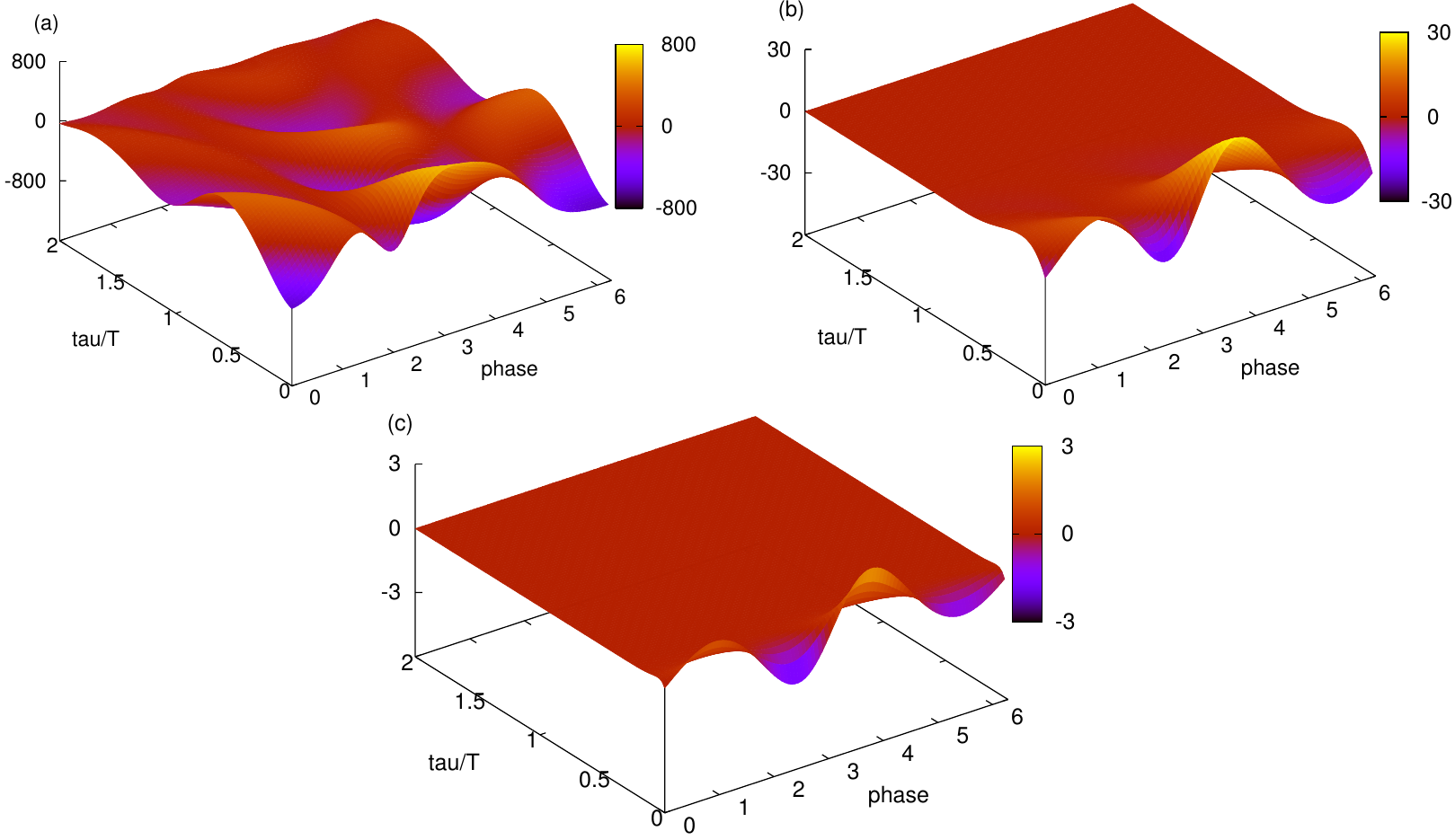}
\caption{(color online) The normalized nonlinear effect of two-pulse action on the Stuart-Landau 
oscillator $\Delta\cdot \e^{-2}$ vs
the phase $\varphi_0$ and the time shift between pulses $\tau$ (normalized by the cycle period), 
for $\e=0.1,\alpha=3$, and three different values of $\mu$ ($\mu=0.1$ in (a), $\mu=0.5$ in (b), and $\mu=2$ in (c)). 
For large $\mu$ the effect is pronounced for very small time intervals 
between two pulses only.}
\label{fig:SLO_1}
\end{figure}

The Stuart-Landau oscillator is a two-dimensional model described in 
polar coordinates as
\[
\dot R=\mu R(1-R^2)\;,\qquad \dot\theta=1+\alpha-\alpha R^2\;.
\]
Here the frequency of the limit cycle, which is a circle with radius $R=1$, 
is normalized to one, parameter $\alpha$ describes nonisochronicity
of oscillations, while $\mu$ is the relaxation rate of the amplitude.  
The phase $\varphi$ defined in the whole plane (except for the origin) is
\[
\varphi=\theta-\frac{\alpha}{\mu}\ln R\;.
\]
Evolution of the variables $R,\theta$ can be explicitly solved as
\[
\begin{pmatrix} R(t)\\ \theta(t)\end{pmatrix}=U^{t-t_0}
\begin{pmatrix} R(t_0)\\ \theta(t_0)\end{pmatrix}=
\begin{pmatrix} [1+\frac{1-R(t_0)^2}{R(t_0)^2}e^{-2\mu(t-t_0)}]^{-1/2}\\ 
\theta(t_0)+t-t_0-\frac{\alpha}{2\mu}\ln(R(t_0)^2+
(1-R(t_0)^2)e^{-2\mu(t-t_0)})\end{pmatrix}\;,
\]
which defines the operator $\mathcal{L}$. We assume that the pulse is 
acting in direction $x$, i.e. at the pulse
\[
R\cos\theta \to R\cos\theta+\epsilon,\qquad R\sin\theta\to R\sin\theta\;.
\]
This fully describes the system, and one can find expressions for the 
PRCs $\Delta\varphi$ and $\delta\varphi$, see Appendix~\ref{ap:prcsl}. 
According to these formulas we calculated the 
nonlinear correction term $\Delta$ and plot it in Fig.~\ref{fig:SLO_1}. 
Here we take $\e_0=\e_1=0.1$
and present results for different $\mu$. As expected, the mostly pronounced 
effect is for small $\mu$.

For this equation it is possible to obtain the leading term in 
order $\sim \e_0\e_1$ in the expansion of the nonlinear
correction term in the pulse strengths analytically (see Appendix~\ref{ap:prcsl}):
 \begin{equation}
\Delta\approx \e_0\e_1 (1+\frac{\alpha^2}{\mu^2})e^{-2\mu\tau}\cos\varphi_0\sin(\varphi_0+\tau)
 \label{eq:sl_error}
 \end{equation}
This expression fits numerics  very good  for $\e\lesssim 0.01$.

\subsection{Example: a modified Stuart-Landau oscillator}
\begin{figure}
\centering
\includegraphics[width=0.9\columnwidth]{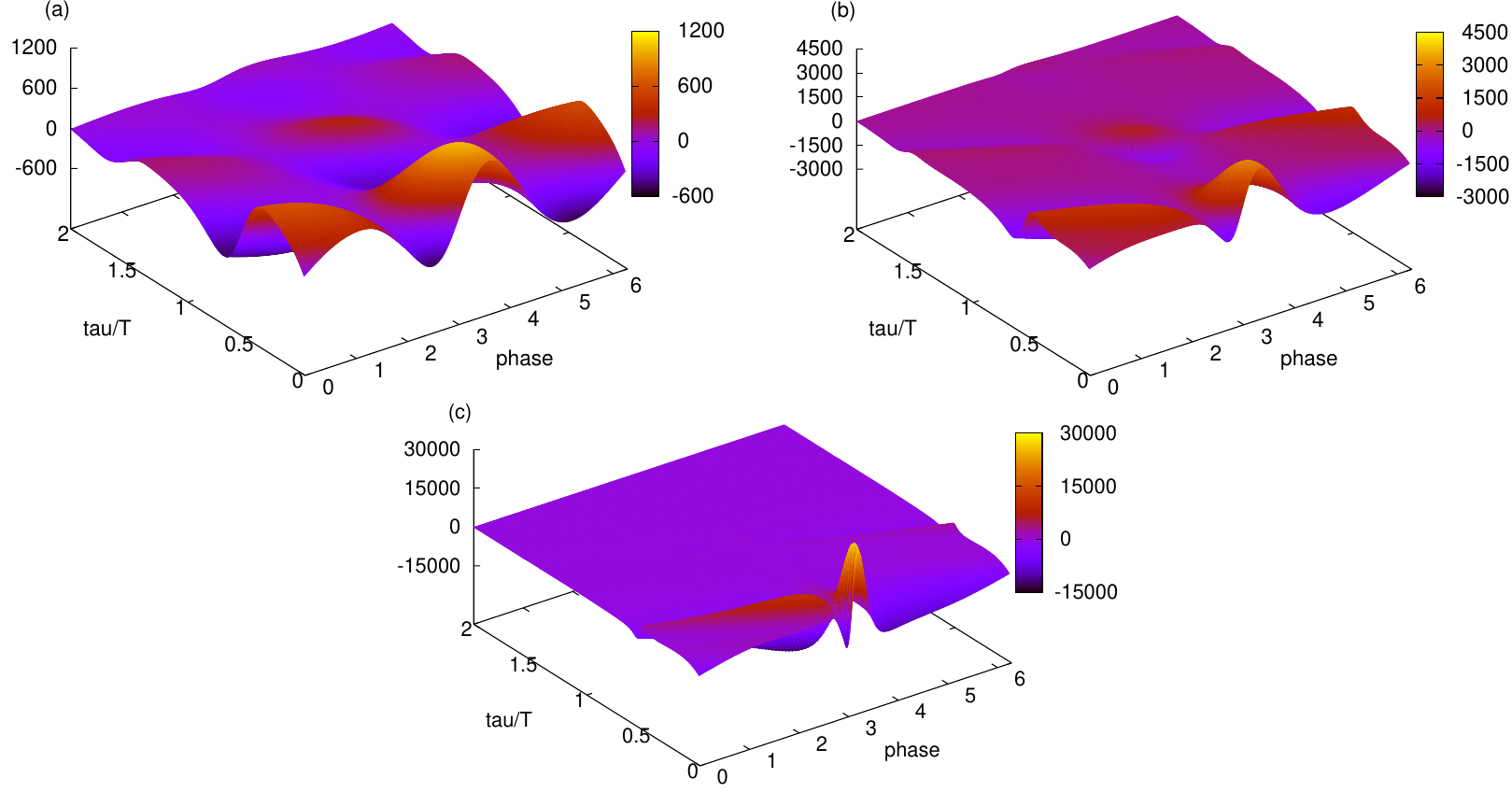}
\caption{(color online) The same as fig.~\ref{fig:SLO_1} but for the modified Stuart-Landau 
oscillator Eq.~(\ref{eq:mslo}). Parameter values: (a) $\alpha=3,\mu=0.1,\e=0.01,b=0.3$; 
(b): $\alpha=3,\mu=0.1,\e=0.01,b=0.7$; (c): $\alpha=3,\mu=0.1,\e=0.001,b=0.95$.}
\label{fig:SLO_2}
\end{figure}

Our second example is a modification of the Stuart-Landau oscillator 
proposed in~\cite{Ermentraut-Glass-Oldeman-12}:
\begin{equation}
\dot R=\mu R(1-R^2)\;,\qquad \dot\theta=1-b r\cos\theta+\alpha-\alpha R^2\;.
\label{eq:mslo}
\end{equation}
Here large values of parameter $b$ produce highly nonuniform growth of 
angle variable $\theta$, so that
the relation between $\varphi$ and $\theta$ is strongly nonlinear. As a 
result, the isochrons crowd
at the region around $\theta\approx 0$ where the evolution of $\theta$ is 
slow. Also the nonlinear correction term
becomes very large at this region, as illustrated in Fig.~\ref{fig:SLO_2}

\clearpage
\section{Neuron Models}

\begin{figure}
\centering
\includegraphics[width=\columnwidth]{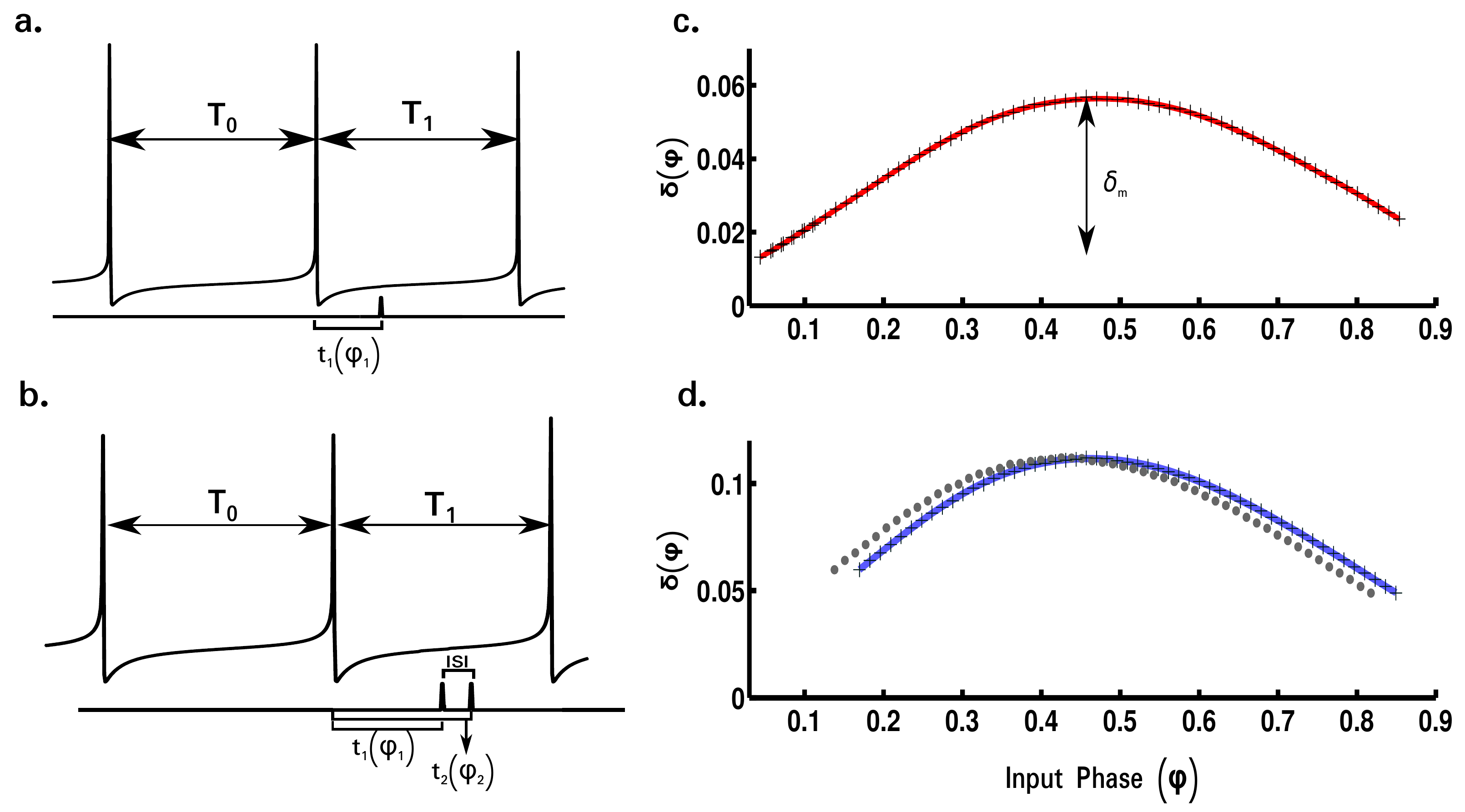}
\caption{(color online) Figure illustrates how PRC is computed for a quadratic integrate fire 
neuron. a. Single pulse, b. Double pulse. $ T_0 $  is the 
original period when no input occurred, $ T_1 $ is  the period when there was 
an input at time $ t_1 $ with input phase  $ \varphi_1 = t_1/T_0 
$. c. PRC $ \delta \varphi  =(T_0-T_1)/T_0 $ for single input (black cross) 
along with its fit (solid red line ($ \bar{\delta}_1(\varphi_1) $)). 
d. PRC for 2 inputs with ISI=10ms.  Gray points indicate the change in period  
against the first input; PRC  with respect to the second input is 
plotted as black cross. The prediction from  superposition principle is plotted 
as solid blue 
line ($ \bar{\delta}_1(\varphi_1) + \bar{\delta}_1(\varphi_2)$).} \label{PRC_measure}
\end{figure}

The PRC is commonly used to describe neuron models. In this context, the PRC  
can characterize the properties of neurons, especially their 
synchronizability.  In many systems a neuron receives inputs from many other 
neurons, therefore, it is critical to understand how multiple 
pulses affect the PRC response of neurons.  Below we test four different 
neuron models for the dual pulse effect. Although generally the theory 
presented in section~\ref{sec:gc} is applicable to spiking neurons as well, 
practically one does not follow the continuous phase of the 
oscillations, but focuses on the spiking events (these events are readily 
available in experiment, too).  Therefore, for spiking neurons, the 
PRC and the non-linear correction term have to be measured in terms of the 
spike times as oppose to phase shifts at arbitrary points as done in 
previous section. It is convenient to normalize the correction term should 
by the peak to trough value of the PRC for single input 
as shown below. We first 
illustrate these definitions in Fig.~\ref{PRC_measure}, using the quadratic 
integrate-and-fire model~\cite{Izhikevich-07}. (Note that 
traditionally  in this context the phase $t/T_0$ and the PRC $ \delta\varphi 
=(T_1-T_0)/T_0$  are normalized by one and not by $2\pi$). This one-
dimensional model corresponds to the pure phase dynamics as in 
section~\ref{sec:ppd}, so it does not demonstrate nonlinear effects of 
deviations from the superposition.  

In general, the models of neurons are classified based on their PRC curves 
as type I and II. The type I PRC,  has only phase advance in response to 
perturbation, while type II PRC includes both phase delay  and 
advance ~\cite{Canavier-06}. We further considered both type I and II neuron 
models in this study. We tested Wang-Buzs\'{a}ki model  
(type I, based on \cite{Wang-Buzsaki-96}), the original Hodgkin-Huxley model 
(type II, based on \cite{Hodgkin52}),  and a modified Hodgkin-Huxley model 
(type I, based on \cite{Mainen-Sejnowski-96}). The equations for  
all the models are given in the Appendix~\ref{app:nm}. All of these models are 
three-dimensional which is required for any deviation from linear superposition. 

\begin{figure}[!h]
\includegraphics[width=\columnwidth]{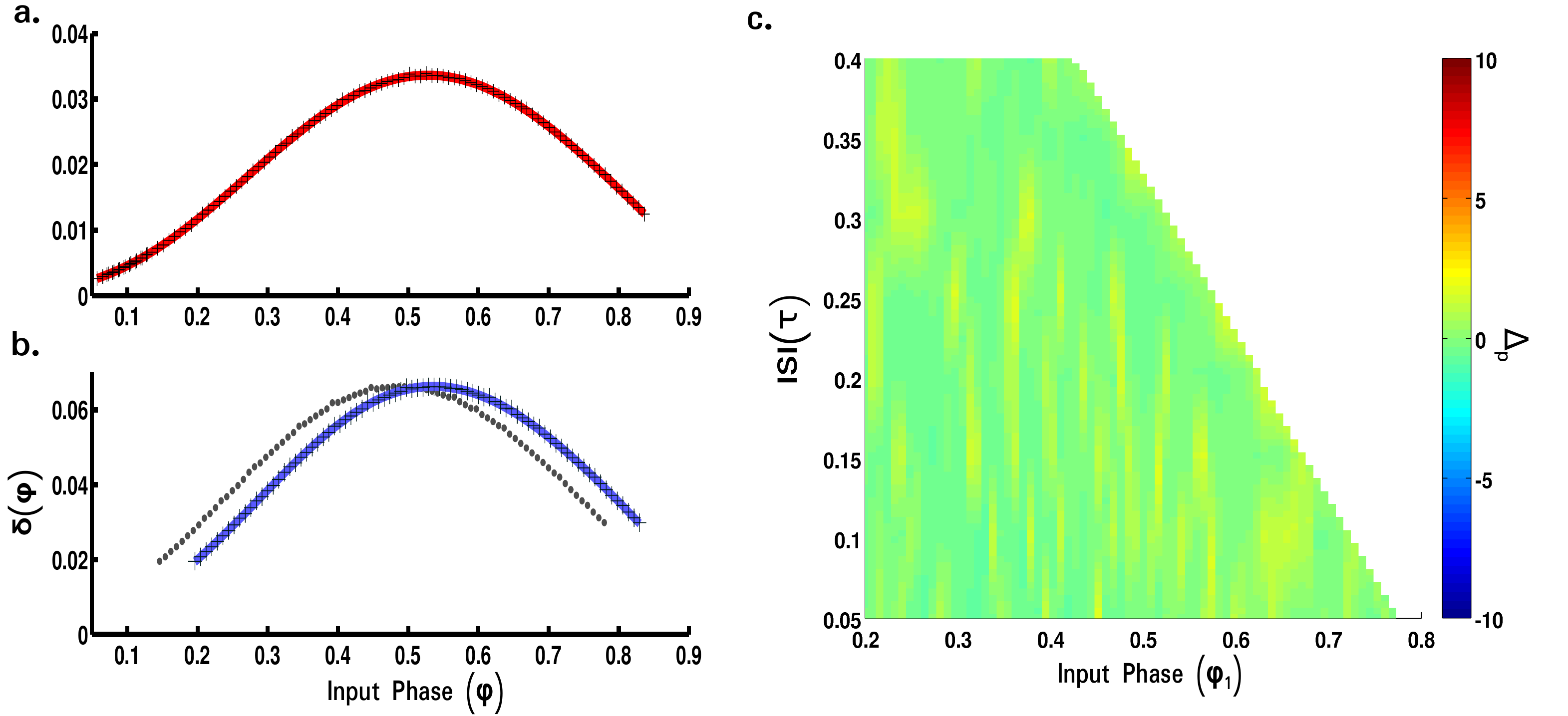}
\caption{(color online) Wang-Buzs\'{a}ki Model. a. PRC for single input along with its 
fit (solid red line) ($ \delta(\varphi) =  (T_0 - T_1)/T_0 $). b. PRC 
for 2 inputs with ISI=10ms. Gray points indicate the PRC with respect 
to the first input and PRC with respect to second input is plotted as 
black cross. The prediction from superposition principle is plotted as 
solid blue line. c. The deviation from linear 
superposition ($ \Delta_p $) for different ISIs. }
\label{PRC_WB_model}
\end{figure}

\begin{figure}[!h]
\includegraphics[width=\columnwidth]{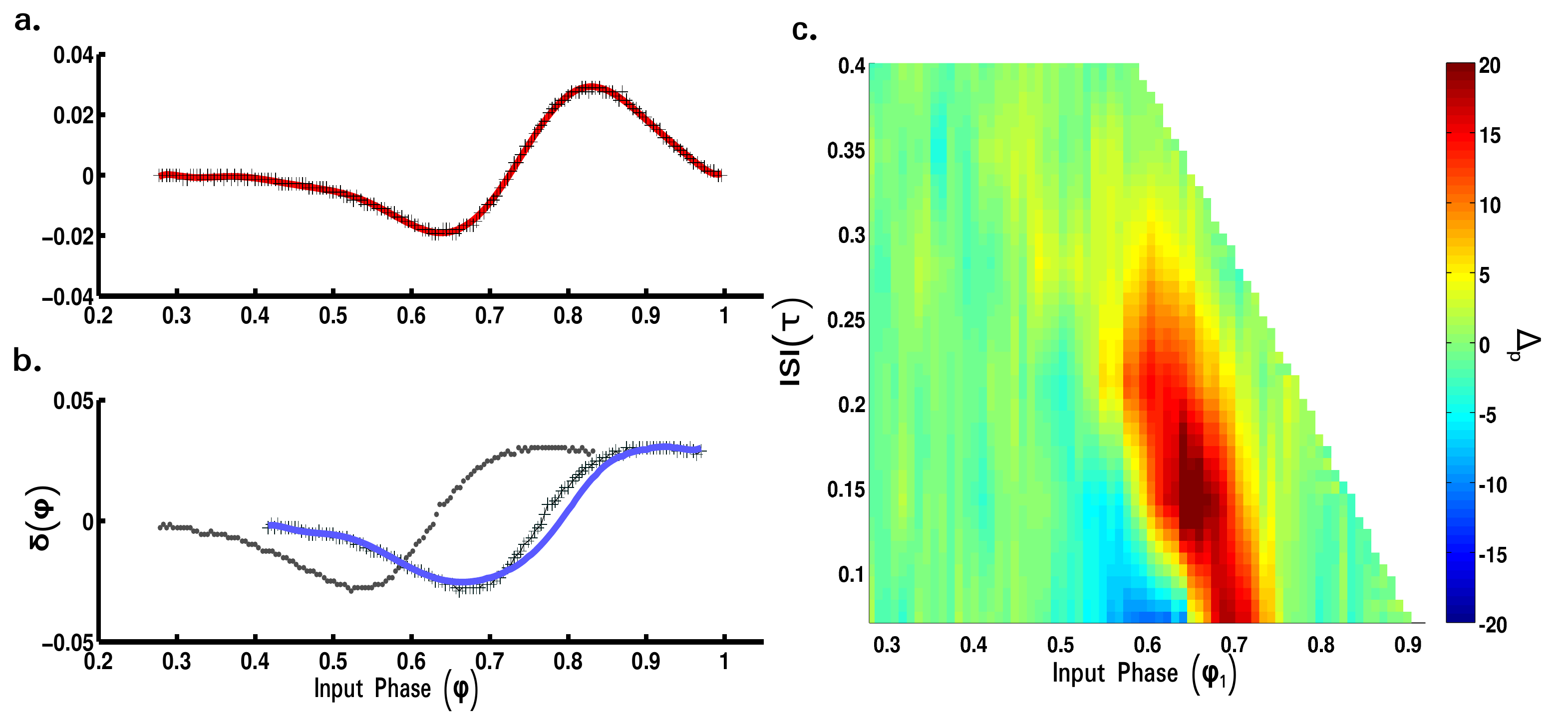}
\caption{(color online) Original Hodgkin-Huxley Model. Legend same as Fig \ref{PRC_WB_model}}
\label{PRC_HH_model}
\end{figure}

\begin{figure}[!h]
\includegraphics[width=\columnwidth]{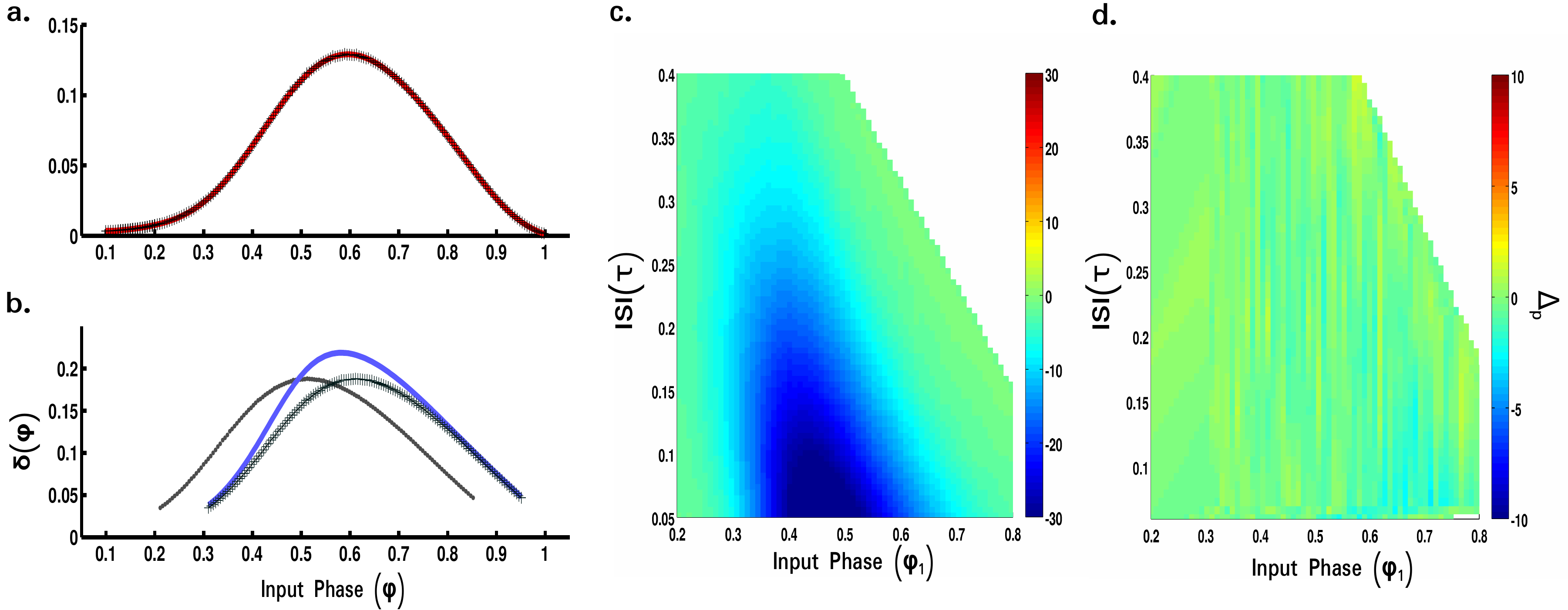}
\caption{(color online) Modified Hodgkin-Huxley Model. a.-c.Legend same 
as Fig \ref{PRC_WB_model} d. Error for modified Hodgin-Huxley model 
with slower evolution, $ \rho $= 0.5 (see Appendix \ref{mHH-appendix} 
for definition of $ \rho $)}
\label{PRC_mHH_model}
\end{figure}

To characterize the deviation from the superposition of two input pulses, 
we use the following quantity
\begin{equation}
 \Delta_p = (\delta_2(\varphi_2) - \bar{\delta}_1(\varphi_1) - \bar{\delta}_1(\varphi_2)) \frac{100}{\delta_m} 
\end{equation}
where, $ \delta_2(\varphi_2) $ is the measured PRC for the second 
pulse, $ \bar{\delta}_1(\varphi_1) $ and $ \bar{\delta}_1(\varphi_2) $ 
are the expected PRCs of single pulses, and $\delta_m$ is the amplitude of the
single-pulse PRC (this normalization provides a better visualization 
of the deviation from the superposition of single-pulse PRCs). We present the 
results for the three neuron models in 
Figs.~\ref{PRC_WB_model}-\ref{PRC_mHH_model}, where the main dependence of 
of $\Delta_p$ on the first input phase and on the interspike interval (ISI)
between two applied pulses $\tau$ is depicted in panel c by a color coding.

We observed that some neuron models show pronounced nonlinear effects, 
as the two-pulse response deviates from the expected PRC based on superposition 
principle, while for other models the linear superposition was able to 
predict the PRC for two-pulses accurately.  The Wang-Buzs\'{a}ki model 
(\ref{PRC_WB_model}) appears to be of the latter type, while both the 
original Hodgkin-Huxley system (\ref{PRC_HH_model}) and its modified 
version (\ref{PRC_mHH_model}) show nonlinear effects in two-pulse PRC.  
The Wang-Buzs\'{a}ki and modified  Hodgkin-Huxley model had type I PRC, 
the original Hodgkin-Huxley model had type II PRC,  suggesting that there 
is no relationship between the type of PRC and the origin of this  deviation. 
Further, the bifurcation type for spiking from the resting state also did not 
determine the existence of error, since the Wang-Buzs\'{a}ki model and the 
modified Hodgkin-Huxley model possess  a saddle-node bifurcation, while the 
original Hodgkin-Huxley model demonstrates an Andronov-Hopf bifurcation. 

\begin{figure}
\includegraphics[width=0.3\columnwidth]{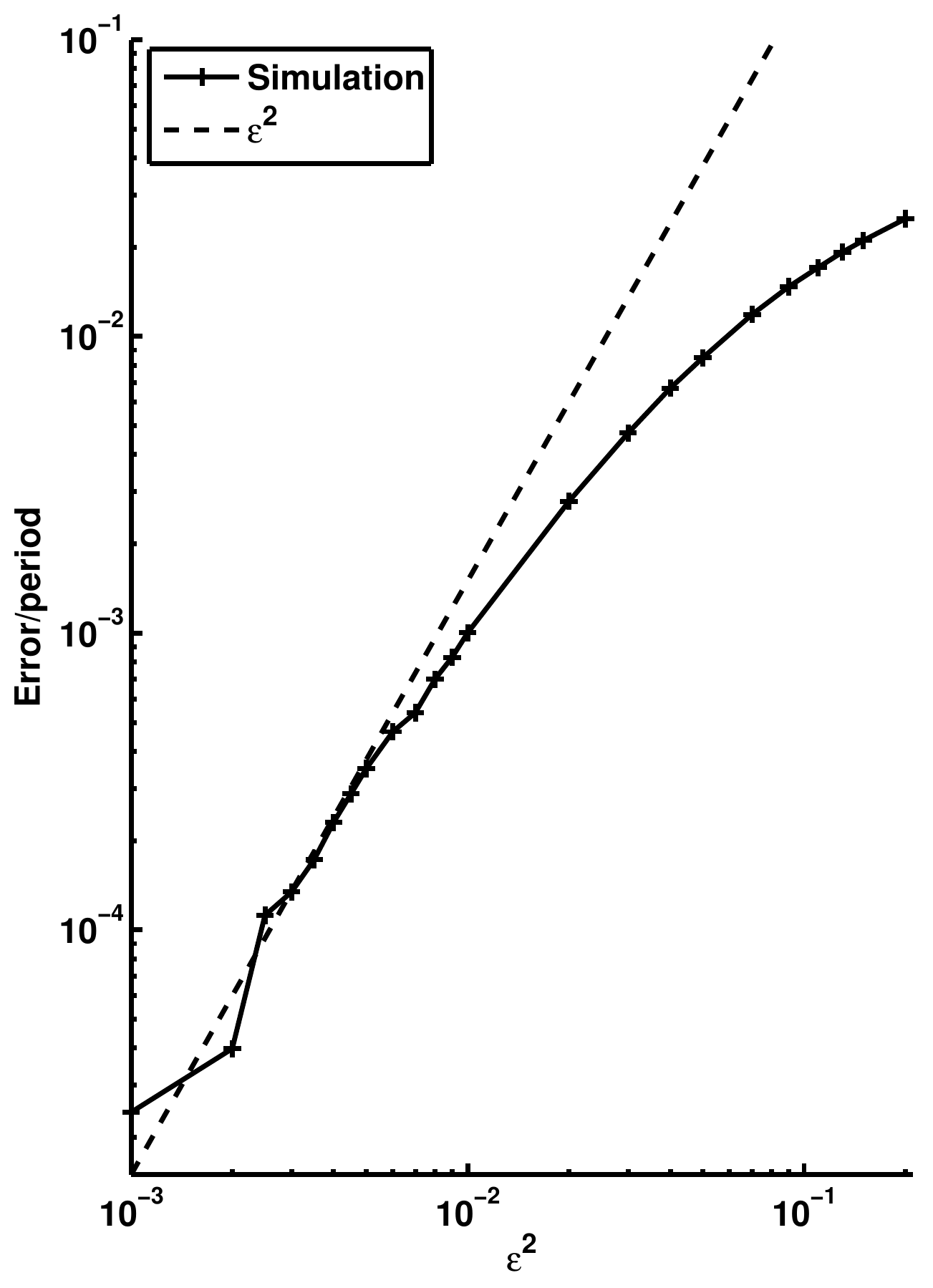}
\caption{Relationship between $ \epsilon^2 $ and $ \Delta $ / period. The 
division by period was used 
to compare with the Landau-Stewart oscillator simulations and does not change 
the relationship 
between $ \epsilon^2 $ and $ \Delta $, since period was same for 
all $ \epsilon^2 $ .}
\label{Quadratic-error}
\end{figure}

There were two main findings from the analytical results of section~\ref{sec:pd} 
that can be tested with the neuronal models presented in this section. The 
first result was that the non-linear correction term was proportional to the 
square  of the perturbation  \ref{eq:sl_error}. We tested this in the modified 
Hodgkin-Huxley model and observed similar qualitative results 
(Fig. \ref{Quadratic-error}). Second, the decay time constant of the 
amplitude term in the oscillator due to perturbation was inversely proportional 
 to the non-linear correction term (\ref{eq:genprc2}). Thus, a slower dynamics 
of the neuron will lead to  lower error. The modified neuron model also showed 
reduced error when the dynamics was  slowed (Fig \ref{PRC_mHH_model}). Thus, 
the main features of  general limit cycle oscillators, and in particular of 
Stuart-Landau oscillator, could be qualitatively extended to some of the 
neuron models.

\section{Conclusion}

In this paper we have developed a theory of a response of an autonomous 
oscillator to two-pulse perturbation. We found that the action of two 
pulses generally deviates from the superposition of two one-pulse 
responses, and this nonlinear effect, which is proportional to the 
product of the perturbations' amplitudes, significantly depends on 
the relation of the interval between the pulses and the relaxation 
time of the oscillator. For fast relaxation and large time interval 
between the pulses, the nonlinear effect vanishes. We have demonstrated 
this property for several models: the standard Stuart-Landau  oscillator, 
the modified version for this oscillator with a highly non-uniform motion 
over the cycle, and for three neuron models, including the classical 
Hodgkin-Huxley system. 

We stress here that in our study, the term ``nonlinearity'' of the
PRC has been used in two contexts. For a one-pulse PRC, nonlinearity means,
that the phase response 
cannot be represented as an amplitude of the pulse multiplied by a 
function of the phase; in particular, the form of the curve may depend
on the pulse amplitude. For two pulses, we use term ``nonlinearity''
to describe a deviation from the superposition principle, this effect in the 
leading order is proportional to the product of the pulses' amplitudes.
We have shown that the nonlinearity of the single-pulse PRC does not necessarily
lead to the the nonlinearity for a two-pulse excitation: the purely phase model
of section \ref{sec:ppd} is a good illustration of this.
Both nonlinear effects may distort a simple picture of the neuron's dynamics
under external forcing.
In particular, in \cite{Izhikevich-07} it was suggested that relatively weak noisy
current to neuron can be applied to obtain the PRC, by solving the equation
that relates the infinitesimal (linear) PRC to the external
voltage through optimization
methods. However, this method does not account for the non-linear
correction which we showed in this study. In some neurons, we showed
that the non-linear effect could have significant effect on the
multi-pulse PRC compared to the single pulse PRC, and thus the continuous
perturbation method may produce erroneous results. Thus, further studies are required
to validate the method proposed in \cite{Izhikevich-07}.

We see two main application fields of our approach. First, 
it can be used for a diagnostics of oscillators. While the 
usual PRC allows to characterize sensitivity of the phase to an 
external action, the nonlinear terms in the two-pulse response 
allow to characterize relaxation processes. In particular, one-pulse  
PRCs of the Wang-Buzs{\'a}ki model and of the modified Hodgkin-Huxley 
model are very similar (cf. Figs.~\ref{PRC_WB_model}a and \ref{PRC_mHH_model}a), 
but their two-pulse PRCs are completely different; this may be useful for 
designing models to fit 
experimental data. Two-pulse PRC can be estimated from experimental data 
and this information can be used to design optimally a model that provides 
a best description of the data. The second field of application is in the 
incorporating these effects in the synchronization theory of pulse coupled 
oscillators. Indeed, in a network an oscillator usually experiences inputs 
from many other units, and in the absence of synchrony the time intervals 
between the incoming pulses can be rather small. In this case the nonlinear 
"interference" of the actions is mostly pronounced, and may contribute 
significantly to the synchronization properties.

\begin{acknowledgments}
G.P.K. would like to thank DAAD for support and University of Potsdam for 
their kind hospitality. This work was supported by NIDCD (grant 1R01DC012943). 

\end{acknowledgments}



%

\appendix

\section{PRC for Stuart-Landau oscillator}
\label{ap:prcsl}
We briefly outline the majors steps in obtaining the approximate non-linear correction term for Stuart-Landau oscillator using an expansion in $\e$. Several intermediate steps are not shown to make presentation brief. Section \ref{sec:slo} provides the phase equation and the evolution operator for Stuart-Landau oscillator. 

At $ t=0 $, let $ R=1 $, $ \theta=\theta_0 $. After $ \e_0 $ pulse at $ t=0 $,  $ \tilde R=\sqrt{(\cos\theta_0+\epsilon_0)^2+\sin^2\theta_0)} $, $ \tilde\theta_0=\arctan(\sin\theta_0/(\cos\theta_0+\epsilon_0)) $, which upon expansion gives, $ \tilde R= 1 + \epsilon_0 \cos\theta_0 + \epsilon_0^2 (\frac{\sin^2\theta_0}{2}) + O(\e^3) $ and $ \tilde\theta_0= \theta_0 - \epsilon_0\sin\theta_0 +  \epsilon_0^2 \sin\theta_0 \cos\theta_0  + O(\e^3) $. From here we ignore all,  $ O(\e^3) $, cubic and higher order terms. \\
\\
The phase shift due to  $ \e_0 $ pulse is,
\begin{gather*}
\Delta \varphi_0=\tilde\theta_0-\theta_0-\frac{\alpha}{\mu} \ln \tilde R=
\arctan(\sin\theta_0/(\cos\theta_0+\epsilon_0))-\theta_0-\frac{\alpha}{\mu} \ln 
\sqrt{(\cos\theta_0+\epsilon_0)^2+\sin^2\theta_0)} \\
\Delta \varphi_0=\arctan\frac{-\epsilon_0\sin\theta_0}{1+\epsilon_0\cos\theta_0} -
\frac{\alpha}{\mu} \ln \sqrt{(\cos\theta_0+\epsilon_0)^2+\sin^2\theta_0)} \\
\approx -\epsilon_0 (\sin\theta_0 + \frac{\alpha}{\mu} \cos\theta_0) + \frac{1}{2} \epsilon_0^2 (\sin 2\theta_0 + \frac{\alpha}{\mu} \cos 2\theta_0)
\end{gather*}

At time $\tau$
\[
\begin{pmatrix} R(\tau)\\ \theta(\tau)\end{pmatrix}=U^{\tau}
\begin{pmatrix} \tilde R\\ \tilde\theta_0\end{pmatrix}
\]

\begin{align*}
\theta(\tau) &= \tilde\theta_0+ \tau - \frac{\alpha}{2\mu} \ln\left(\tilde R^2+(1-\tilde R^2)\exp(-2\mu\tau)\right) \\
&\approx \tau + \theta_0 - \epsilon_0 \sin\theta_0 + \epsilon_0^2 \sin\theta_0 \cos\theta_0 - \frac{\alpha}{2\mu} \ln \left( (1+2\epsilon_0\cos\theta_0 + \epsilon_0^2) + (-2\epsilon_0 \cos\theta_0 - \epsilon_0^2) e^{-2\mu\tau} \right) \\
&\approx \tau + \theta_0 - \epsilon_0 [ \sin\theta_0 + \frac{\alpha}{\mu} \cos (1 - e^{-2\mu\tau})] + \epsilon_0^2 [ \sin\theta_0 \cos\theta_0 - \frac{\alpha}{2\mu} ( 1 - e^{-2\mu\tau} - 2 \cos^2\theta_0 (1 - e^{-2\mu\tau})^2 )] \\
&\equiv \tau + \theta_0 - \epsilon_0 E + \epsilon_0^2 F\\
R(\tau)&= \big[ \frac{\tilde R^2}{\tilde R^2+(1-\tilde R^2)e^{-2\mu\tau}} \big]^{1/2} 
\approx \big[ \frac{1+2\epsilon_0\cos\theta_0+\epsilon_0^2}{1+2\epsilon_0\cos\theta_0+\epsilon_0^2+(-2\epsilon_0\cos\theta_0-\epsilon_0^2)  e^{-2\mu\tau}} \big]^{1/2}\\
R(\tau)^2 &\approx \frac{1+2\epsilon_0\cos\theta_0+\epsilon_0^2}{1+2\epsilon_0\cos\theta_0( 1 - e^{-2\mu\tau} ) + \epsilon_0^2 ( 1 - e^{-2\mu\tau}) }\\
&\approx 1 + 2 \epsilon_0\cos\theta_0 e^{-2\mu\tau} - 4 \epsilon_0^2 \cos^2\theta_0 ( 1 - e^{-2\mu\tau}) + \epsilon_0^2 e^{-2\mu\tau} + 4 \epsilon_0^2 \cos^2\theta_0 ( 1 - e^{-2\mu\tau})^2 \\
&\approx 1 + 2 \epsilon_0\cos\theta_0 e^{-2\mu\tau} + \epsilon_0^2 [- 4 \cos^2\theta_0 ( 1 - e^{-2\mu\tau})e^{-2\mu\tau} + e^{-2\mu\tau} ]\\
&\equiv 1 + \epsilon_0 C + \epsilon_0^2 D
\end{align*}

After the second pulse ($ \e_1 $) at time $ \tau $, $ \hat R(\tau)=\sqrt{(R(\tau)\cos\theta(\tau)+\epsilon_1)^2+ (R(\tau)\sin\theta(\tau))^2}  $ and 
$ \hat\theta(\tau)=\arctan(R(\tau)\sin\theta(\tau)/(R(\tau)\cos\theta(\tau)+ \epsilon_1)) $. 
So the final phase shift is, $ \Delta \varphi_{0,1}=\hat\theta(\tau)-\theta_0-\tau-\frac{\alpha}{2\mu}\ln \hat R^2(\tau) $, which needs to be compared to prediction from linear superposition which is given as $ \Delta \varphi_0 + \Delta\varphi_1 $. We first expand $ \cos\theta(\tau) $, $ \hat R(\tau) $, $ \hat \theta(\tau) $, since it is used in the comparison as,
\begin{align*}
\cos\theta(\tau) &= \cos(\tau + \theta_0 - \epsilon_0 E + \epsilon_0^2 F)\\
&\approx \cos(\tau + \theta_0)- \epsilon_0 E \sin(\tau + \theta_0) + \epsilon_0^2 (-\frac{1}{2} E^2 \cos(\tau + \theta_0) - F \sin(\tau + \theta_0) )\\
\hat R^2(\tau) &= R^2(\tau) + 2 R(\tau) \epsilon_1 \cos\theta(\tau) + \epsilon_1^2\\
&\approx 1 + \e_0 C + \e_0^2 D + 2\e_1(1+\frac{1}{2} \e_0 C) (\cos(\theta_0 +\tau) + \e_0 E \sin(\tau + \theta_0) ) + \epsilon_1^2\\
\hat \theta(\tau) &= \arctan( R(\tau) \sin\theta(\tau)/ (R(\tau)\cos\theta(\tau) + \e_1)) \\
&\approx \theta(\tau) - \frac{\e_1}{R(\tau)} \sin\theta(\tau) + \frac{\e^2}{R^2(\tau)} \sin\theta(\tau)\cos\theta(\tau)\\
&\approx \tau + \theta_0 + \epsilon_0 E - \e_1 \sin(\tau + \theta_0) + \epsilon_0^2 F +\\
& \e_0\e_1[-\cos(\theta_0 +\tau) E + \frac{1}{2} \sin(\theta_0 +\tau) C] + \e_1^2 \sin(\theta_0 +\tau) \cos(\theta_0 +\tau)
\end{align*}

\begin{align*}
\Delta \varphi_0 + \Delta \varphi_1 &= \Delta \varphi_0 + \arctan\left(\frac{\sin(\theta_0+\Delta\varphi_0+\tau)}{(\cos(\theta_0+\Delta\varphi_0+\tau)+\epsilon_1)}\right) -(\theta_0+\Delta\varphi_0+\tau)\\
&- \frac{\alpha}{2\mu} \ln (\cos(\theta_0+\Delta\varphi_0+\tau)+\epsilon_1)^2+ \sin^2(\theta_0+\Delta\varphi_0+\tau)\\
&\approx \Delta \varphi_0 + \arctan\frac{-\epsilon_1\sin(\theta_0 + \Delta\varphi_0 + \tau)}{1+\epsilon_1\cos(\theta_0+\Delta\varphi_0+\tau)}-\frac{\alpha}{\mu} \ln 
(1 + 2 \e_1 \cos(\theta_0 + \Delta\varphi_0 + \tau) + \e_1^2)\\
&\approx -\epsilon_0 (\sin\theta_0 + \frac{\alpha}{\mu} \cos\theta_0) + \frac{1}{2} \epsilon_0^2 (\sin 2\theta_0 + \frac{\alpha}{\mu} \cos 2\theta_0) \\
& + \e_1[\sin(\theta_0 + \Delta\varphi_0 + \tau) + \frac{\alpha}{\mu} \cos(\theta_0+\Delta\varphi_0+\tau)] \\
& + \frac{1}{2} \e_1^2 [\sin2(\theta_0 + \Delta\varphi_0 + \tau) + \frac{\alpha}{\mu} \cos2(\theta_0 + \Delta\varphi_0 + \tau)]\\
\Delta \varphi_{0,1} &=\hat\theta(\tau) - \theta_0 - \tau - \frac{\alpha}{2\mu}\ln \hat R^2(\tau)\\
&= \epsilon_0 E - \e_1 \sin(\tau + \theta_0) + \epsilon_0^2 F + \e_0\e_1[-\cos(\theta_0 +\tau) E + \frac{1}{2} \sin(\theta_0 +\tau) C] \\
&+ \e_1^2 \sin(\theta_0 +\tau) \cos(\theta_0 +\tau) \\
&- \frac{\alpha}{2\mu}\ln [1 + \e_0 C + \e_0^2 D + 2\e_1(1+\frac{1}{2} \e_0 C) (\cos(\theta_0 +\tau) + \e_0 E \sin(\tau + \theta_0) ) + \epsilon_1^2]
\end{align*}

Finally, the difference between the prediction from superposition and actual phase reset (after substitutions) is given as, 
\begin{align*}
& \Delta \varphi_{0,1} - (\Delta \varphi_0 + \Delta \varphi_1) = \\
& \epsilon_0 [ \sin\theta_0 + \frac{\alpha}{\mu} \cos (1 - e^{-2\mu\tau})] - \e_1 \sin(\tau + \theta_0) + \epsilon_0^2  [ \sin\theta_0 \cos\theta_0 - \frac{\alpha}{2\mu} ( 1 - e^{-2\mu\tau} - 2 \cos^2\theta_0 (1 - e^{-2\mu\tau})^2 )] \\
&+ \e_0\e_1\big[-\cos(\theta_0 +\tau) [ \sin\theta_0 + \frac{\alpha}{\mu} \cos (1 - e^{-2\mu\tau})] + \frac{1}{2} \sin(\theta_0 +\tau) 2\cos\theta_0 e^{-2\mu\tau} \big] 
+ \e_1^2 \sin(\theta_0 +\tau) \cos(\theta_0 +\tau) \\
&- \frac{\alpha}{2\mu}\ln \Big[ 1 + \e_0 2\cos\theta_0 e^{-2\mu\tau} + \e_0^2 [- 4 \cos^2\theta_0 ( 1 - e^{-2\mu\tau})e^{-2\mu\tau} + e^{-2\mu\tau} ] \\
&\qquad\qquad + 2\e_1(1+\frac{1}{2} \e_0 2\cos\theta_0 e^{-2\mu\tau} ) (\cos(\theta_0 +\tau) + \e_0 [ \sin\theta_0 + \frac{\alpha}{\mu} \cos (1 - e^{-2\mu\tau})] \sin(\tau + \theta_0) ) + \epsilon_1^2 \Big]\\
& -\Big[ -\epsilon_0 (\sin\theta_0 + \frac{\alpha}{\mu} \cos\theta_0) + \frac{1}{2} \epsilon_0^2(\sin 2\theta_0 + \frac{\alpha}{\mu} \cos 2\theta_0) + \e_1[\sin(\theta_0 + \Delta\varphi_0 + \tau) \\ 
&+ \frac{\alpha}{\mu} \cos(\theta_0+\Delta\varphi_0+\tau)] + \frac{1}{2} \e_1^2 [\sin2(\theta_0 + \Delta\varphi_0 + \tau) + \frac{\alpha}{\mu} \cos2(\theta_0 + \Delta\varphi_0 + \tau)] \Big]\\
\end{align*}
which upon several steps of algebraic reductions and ignoring cubic or higher $ \e $ terms gives,
\[
\Delta \varphi_{0,1} - (\Delta \varphi_0 + \Delta \varphi_1) \approx \e_0\e_1 (1+\frac{\alpha^2}{\mu^2})e^{-2\mu\tau}\cos\varphi_0\sin(\varphi_0+\tau)
\]

\section{Neuron models}
\label{app:nm}
In these models $ I_{stim} $ is the external input. 

Wang-Buz\'{a}ki Model~\cite{Wang-Buzsaki-96}:
\begin{equation}
  \begin{aligned}
    \dot v &= -0.1(v + 65) - 9 n^4 (v + 90) - 35 m_\infty^3 h (v - 55) -I_{Stim} \\
    \dot h & = (h_\infty-h)/h_\tau \qquad  \dot n = (n_\infty-n)/n_\tau \qquad  m_\infty = \alpha_m/(\alpha_m+\beta_m) \\
    \alpha_m &= -(v+35)/(10 (e^{-(v+35)/10}-1)),  \qquad   \beta_m =  4 e^{-(v+60)/18} \\ 
    h_\infty &= \alpha_h/(\alpha_h+\beta_h), \qquad h_\tau = 1/(5(\alpha_h+\beta_h)) \qquad  \alpha_h = .07 e^{-(v+58)/20}, \qquad   \beta_h = 1 / (e^{-(v+28)/10} + 1) \\
    n_\infty &= \alpha_n/(\alpha_n+\beta_n),  \qquad  n_\tau = 1/(5(\alpha_n+\beta_n)) \\
    \alpha_n &= -.01(v+34)/(e^{-(v+34)/10}-1) \quad  \beta_n = .125 e^{-(v+44)/80} \\
  \end{aligned}
\end{equation}

Hodgkin-Huxley Model~\cite{Hodgkin52}:
\begin{equation}
  \begin{aligned}
    \dot v &= -0.5 (v + 65) - 36 n^4(v + 77) -120  m_\infty^3h(v - 50) - I_{Stim}\\
    \dot h &= (h_\infty-h)/h_\tau  \qquad  \dot n = (n_\infty-n)/n_\tau \qquad  m_\infty = \alpha_m/(\alpha_m+\beta_m) \\
    \alpha_m &= -0.1(v+40)/(e^{-(v+40)/10}-1)  \qquad  \beta_m =  4 e^{-(v+65)/18} \\   
    h_\infty &= \alpha_h/(\alpha_h+\beta_h)  \qquad h_\tau = 1/(\alpha_h+\beta_h) \qquad  \alpha_h = .07 e^{-(v+65)/20} \qquad  \beta_h = 1 / (e^{-(v+35)/10} + 1) \\
    n_\infty &= \alpha_n/(\alpha_n+\beta_h)  \qquad n_\tau = 1/(\alpha_n+\beta_n) \\
    \alpha_n &= -.01(v+55)/(e^{-(v+55)/10}-1) \qquad \beta_n = .125 e^{-(v+65)/80} \\
  \end{aligned}
\end{equation}

Modified Hodgkin-Huxley Model~\cite{Mainen-Sejnowski-96}:
\label{mHH-appendix}
\begin{equation}
  \begin{aligned}
    \dot v &= (-0.0317 (v + 77.8) - 30.032  m_\infty^3 h (v - 49.8) - 5.315 n (v + 100.4) -I_{Stim})/0.75 \\
    \dot h &= - \rho (h - h_\infty)/h_\tau \qquad  \dot n = - \delta (n - n_\infty)/n_\tau \qquad   m_\infty = \alpha_m/(\alpha_m+\beta_m) \\ 
    \alpha_m &= 0.182 (v+25)/(1-e^{-(v+25)/9}) \qquad  \beta_m = 0.124 (-v-25)/(1-e^{-(-v-25)/9})\\  
    h_\infty &= \alpha_h/(\alpha_h+\beta_h)  \qquad h_\tau = 1/(2.953 (\alpha_h+\beta_h))\\
    h_\infty &= 1/(1+e^{(v+55)/6.2}) \\
    \alpha_h &= 0.024 (v+40)/(1-e^{-(v+40)/5});  \beta_h = 0.0091 (-v-65))/(1-e^{-(-v-65)/5}) \\
    n_\infty &= \alpha_n/(\alpha_n+\beta_n)  \qquad n_\tau = 1/(2.953 (\alpha_n+\beta_n))  \\
    \alpha_n &= 0.02*(v-25)/(1-e^{-(v-25)/9})  \qquad   \beta_n = -0.002*(v-25)/(1-e^{(v-25)/9})\\  
  \end{aligned}
\end{equation}
where, $ \rho $ is the factor which was reduced to 0.5 in Fig \ref{PRC_mHH_model}.

\end{document}